\documentclass[12pt,epsf]{article}
\usepackage{amssymb,amsmath,amsbsy}
\usepackage{graphicx}
\usepackage{setspace}
\usepackage{subfigure}
\usepackage{cite} 
\newcommand{\be}{\begin{equation}}
\newcommand{\ee}{\end{equation}}
\newcommand{\bea}{\begin{eqnarray}}
\newcommand{\eea}{\end{eqnarray}}
\newcommand{\bear}{\begin{eqnarray}}
\newcommand{\eear}{\end{eqnarray}}
\newcommand{\beas}{\begin{eqnarray*}}

\newcommand{\eeas}{\end{eqnarray*}}
\newcommand{\ba}{\begin{array}}
\newcommand{\ea}{\end{array}}

\newcommand{\ra}

\newcommand{\pd}[2][1]{\ifnum#1=1 \frac{\partial}{\partial {#2}} \else
  \frac{\partial^#1}{\partial {#2}^{#1}}\fi}
\newcommand{\dpd}[2][1]{\ifnum#1=1 \dfrac{\partial}{\partial {#2}} \else
  \frac{\partial^#1}{\partial {#2}^{#1}}\fi}
\newcommand{\td}[2][1]{\ifnum#1=1 \frac{d}{d{#2}} \else
  \frac{d^#1}{d{#2}^{#1}}\fi}

\newcommand{\nbox}{{\,\lower0.9pt\vbox{\hrule \hbox{\vrule height 0.2 cm \hskip 0.19 cm \vrule height 0.2 cm}\hrule}\,}}

\def\href#1#2{#2}

\usepackage{color}

\usepackage{xcolor}
\usepackage[citebordercolor=green, linkbordercolor={ 0 0 1}, linktocpage=true]{hyperref}

\begin{document}
\begin{titlepage}
\begin{NoHyper}
\hfill
\vbox{
    \halign{#\hfil         \cr
           } 
      }  
\vspace*{20mm}
\begin{center}
{\Large \bf Holographic Consequences of a \\ 
No Transmission Principle}

\vspace*{15mm}
\vspace*{1mm}
Netta Engelhardt and Gary T. Horowitz
\vspace*{1cm}
\let\thefootnote\relax\footnote{engeln@physics.ucsb.edu, gary@physics.ucsb.edu}

{Department of Physics, University of California\\
Santa Barbara, CA 93106, USA}

\vspace*{1cm}
\end{center}
\begin{abstract}
Two quantum field theories whose Hilbert spaces do not overlap cannot transmit
 a signal to one another. From this simple principle, we deduce some highly nontrivial consequences for holographic quantum gravity. These include: (1) certain cosmological bounces are forbidden, (2) generic singularities inside black holes cannot be resolved, and (3) traversable wormholes do not exist.   At the classical level, this principle rules out certain types of naked singularities and suggests that new singularity theorems should exist. 

\end{abstract}
\end{NoHyper}

\end{titlepage}
\tableofcontents
\vskip 1cm
\begin{spacing}{1.2}
\section{Introduction}\label{sec:intro}

A true understanding of singularities, both in cosmology and in the black hole interior, requires a nonperturbative formulation of a quantum theory of gravity. Perhaps the best-understood approach to quantum gravity is gauge/gravity duality, which relates quantum string theory on asymptotically locally Anti-de Sitter (AdS) backgrounds -- the bulk -- to quantum field theories\footnote{Note that we do not restrict to conformal field theories, as gauge/gravity duality can also apply to theories without conformal invariance, provided that these theories have a UV fixed point.} living on the conformal boundary of the bulk~\cite{Mal97, Wit98a, GubKle98}. At its strongest form, the duality provides a definition of nonperturbative quantum gravity in terms of a field theory. A powerful feature of this duality is that statements that are hard to establish on one side of the duality are often much easier to prove or derive on the dual counterpart. Unitarity of the evolution is an early example of a simple statement about field theory with a highly nontrivial consequence for black hole evaporation in the gravitational dual. This is the direction that we pursue in this paper: deducing significant implications for quantum gravity from a simple field theory principle.

The principle we employ is intuitive:    
a signal cannot be transmitted between two quantum field theories defined on separate Hilbert spaces. We call this the \textit{No Transmission Principle}. In general, two quantum field theories have separate Hilbert spaces when they are defined on separate spacetimes. When the field theories in question are conformally invariant, there is a subtlety: two spacetimes which a priori appear disconnected can possibly be conformally mapped to one connected spacetime. Two disjoint copies of a conformal field theory (CFT) on $d$-dimensional Minkowski spacetime, for example, can be mapped into disjoint finite subsets of the same CFT on the Einstein static universe. The two original copies of the CFT are then defined on the same Hilbert space, and 
signals can then propagate from one to the other through the CFT on the Einstein static universe.  In the much more common situation in which two CFTs cannot be conformally embedded in a larger CFT in this sense, their Hilbert spaces cannot overlap, and therefore no 
signals may be exchanged between the two (see Fig.~\ref{fig:Einstein} for an illustration).

The application to holography is immediate. If two field theories with holographic duals do not share a Hilbert space, then no 
signal can be transmitted between their bulk duals. This is reminiscent of the Gao--Wald theorem~\cite{GaoWal00}: signals between two boundary points travel faster on the boundary than in the bulk; here we say that 
signals can only propagate through the bulk if it can do so on the boundary. This results in nontrivial constraints on the structure of bulk spacetimes with cosmological singularities or black holes. We note that while certain black hole spacetimes might appear to present a violation of this line of reasoning (e.g. Reissner-Nordstrom-AdS) by allowing 
signals to propagate from one asymptotic boundary to another through the inner horizon, the instability of the inner horizon guarantees that no actual violations occur. This is discussed at length in Sec~\ref{sec:nonsingular}.

The power of the No Transmission Principle in holography is in its broad applicability to quantum string theory in \textit{any regime}. The statement of the principle omits any mention of $N$ or the 't Hooft coupling, so any constraints obtained from this principle are valid when the bulk theory is anything from classical Einstein gravity to full quantum string theory.  At the classical level, it implies new results about asymptotically AdS solutions of general relativity, which may presumably be proved using geometry alone. At the quantum level, it yields new insights into quantum gravity, which cannot be obtained directly from our current understanding of the theory.

We organize our investigation into two different categories based on the extent of evolution of the dual field theory. In particular, we separately consider field theories in which evolution stops at some finite time -- ``singular'' field theories -- and ones in which evolution continues indefinitely, which we term ``nonsingular'' (in a conformally--invariant sense to be defined in Sec.~\ref{sec:NTP}). Application of the No Transmission Principle to two singular field theories yields constraints on the behavior and resolution of cosmological singularities; the implications for two nonsingular field theories place limits on black hole singularity resolution and wormholes. Below we provide a heuristic discussion of the principle's implications separately for singular and nonsingular CFTs.

\noindent \textbf{Singular Field Theories:} The No Transmission Principle yields a simple result when applied to singular field theories: if the boundary constituents are two singular field theories, then their corresponding Hilbert spaces have no mutual overlap. In this case, the field theory singularity must extend through the bulk, cutting off bulk evolution at finite time; the alternative would allow 
signals to propagate between the field theories through the bulk even though such 
propagation is forbidden  due to the No Transmission Principle.

We may now investigate consequences of terminated bulk evolution in various regimes of quantum string theory. First, at the level of classical General Relativity (large $N$ and large coupling), we obtain nontrivial statements about the structure of singular solutions to general relativity: a cosmological singularity on the boundary results in a bulk singularity which extends to infinity and cannot end in the interior, leaving a smooth region behind. This agrees with intuition, as a spacetime which has a singularity with a ``hole'' would constitute a gross violation of cosmic censorship as well as global hyperbolicity.  Such a singularity would arise from smooth initial data in the past and an observer traveling through the hole would be able to see the singularity on the other side. There are various cosmological singularity theorems in general relativity which have a similar conclusion, but those assume the  strong energy condition, which is violated by reasonable and more pertinently extant matter such as massive scalar fields. 

At the level of classical string theory (infinite $N$ and finite coupling), we find constraints on stringy singularities. Certain types of cosmological singularities can be smoothed out by closed string tachyon condensation \cite{McGSil05}. In the context of holography, the No Transmission Principle shows that a tachyon condensate cannot mediate signal transmission from a collapsing spacetime to an expanding one.

The application of this argument  to full quantum gravity is of greatest interest. Gauge/gravity duality at finite $N$ and finite coupling still has a fixed (classical) metric on the conformal boundary, but the metric in the interior is subject to arbitrarily large quantum and stringy fluctuations.  The No Transmission Principle certainly applies to field theories at finite $N$ and coupling, leading to the conclusion that holographic quantum gravity forbids a large class of bounces through cosmological singularities\footnote{We are only able to rule out holographic bounces for which the dual field theory is genuinely singular. We are  aware of only one example of a cosmological singularity in holography for which the dual field theory  may be nonsingular in the sense of Sec.~\ref{sec:singular}. See Sec.~\ref{sec:bounce} for discussion.}.

Cosmological bounces have been extensively investigated in prior literature, ranging from classical solutions which violate the strong energy condition~\cite{GraHor11,BhaCha15}, to bounces in nonholographic  theories of quantum gravity, and even bouncing cosmologies in holography. With regard to the first, we are not aware of any such classical solutions with asymptotically AdS initial data that  would contradict our results. Our analysis does not apply to the second category, which includes  loop quantum gravity \cite{Boj08, AshSin11}, as well as string inspired models that use T-duality \cite{BraVaf88, GasVen02} or colliding branes \cite{Leh08}.   Our conclusions for holographic bounces could only be avoided in the following way: by matching the state of one dual field theory prior to the singularity to the state of another dual theory after the singularity \cite{CraHer07, SmoTur12}. This extra input is not part of standard holography and, in our opinion, is unnatural and  not justified. Furthermore, there are indications that a suitable rule does not exist. We will discuss this further in Sec.~\ref{sec:discussion}. \\

\noindent \textbf{Nonsingular Field Theories:} The No Transmission Principle may be similarly applied to nonsingular field theories to yield fruitful results constraining the black hole interior and forbidding traversable wormholes that connect two asymptotically AdS regions of spacetime. The line of reasoning is similar to the cosmological case: two field theories on separate spacetimes which have maximal evolution -- i.e. cannot be evolved any further in any conformal frame -- must have non-overlapping Hilbert spaces. They therefore cannot transmit
signals to one another.

The above argument immediately shows that quantum string theory forbids the resolution of a  Schwarzschild black hole future singularity into a white hole singularity of another asymptotic region. That would allow signals to propagate from one asymptotically AdS bulk to another, inducing forbidden transmission between nonsingular field theories. More generally, a charged or rotating AdS black hole has an inner horizon which is known to be classically unstable~\cite{SimPen73, PoiIsr90}: generic perturbations will 
turn this horizon into a weak null singularity \cite{Ori91, Daf04} (see Sec.~\ref{sec:nonsingular} for more detail). The No Transmission Principle implies that there is no evolution past this singularity into another asymptotic region in holographic quantum gravity. Finally, the same line of reasoning shows that traversable wormholes between two asymptotically AdS bulks are forbidden by quantum string theory, as such a setup would allow signals to propagate from one asymptotically AdS region to another.\\

In the next section we introduce the No Transmission Principle in greater detail. Singular and nonsingular field theories are defined in Sec.~\ref{sec:singular}, where the No Transmission Principle is used to derive constraints on cosmological bounces. Sec.~\ref{sec:nonsingular} contains applications to nonsingular field theories, and Sec.~\ref{sec:discussion} contains some concluding remarks. This includes a brief discussion of applications of the No Transmission Principle to superselection sectors in holography~\cite{MarWal12}.

\section{No Transmission Principle}\label{sec:NTP}

When can 
signals be exchanged between two quantum field theories? Excitations in one field theory can only make sense in the other if the operator algebra of each field theory acts on a common Hilbert space\footnote{We thank D. Jafferis for suggesting this formulation.}.  
More precisely, consider two field theories on fixed classical, non-overlapping geometries\footnote{We will not consider backreaction, and for our purposes these spacetimes need not  be dynamical.} of the same dimensionality with Hilbert spaces $\mathcal{H}_{1}$ and $\mathcal{H}_{2}$, respectively. We will say that the two field theories are \textit{independent} if the joint Hilbert space of the system can be written as a tensor product:   $\mathcal{H}=\mathcal{H}_{1}\otimes \mathcal{H}_{2}$, and when acting on $\mathcal{H}$, every operator of the first theory takes the form $\Phi_1 \otimes I_2$, and every operator of the second theory takes the form $I_1\otimes \Phi_2$, where $I_i$ is the identity operator on $\mathcal H_i$. Otherwise, the two field theories are \textit{dependent}.  \\

\noindent \textbf{No Transmission Principle (NTP):} {\it Two independent quantum field theories cannot transmit 
signals to one another.}\\

At first sight, the simplicity of this statement may belie its utility. The power of the principle stems from its applicability to gauge/gravity duality without reference to the coupling or the number of degrees of freedom of the field theory. In the holographic context, it states that if two holographic field theories cannot influence one another on the boundary, then they cannot do so through the bulk, be it classical, stringy, or quantum.

\begin{figure}[t]
\centering
\subfigure[]{
\includegraphics[height= 9 cm]{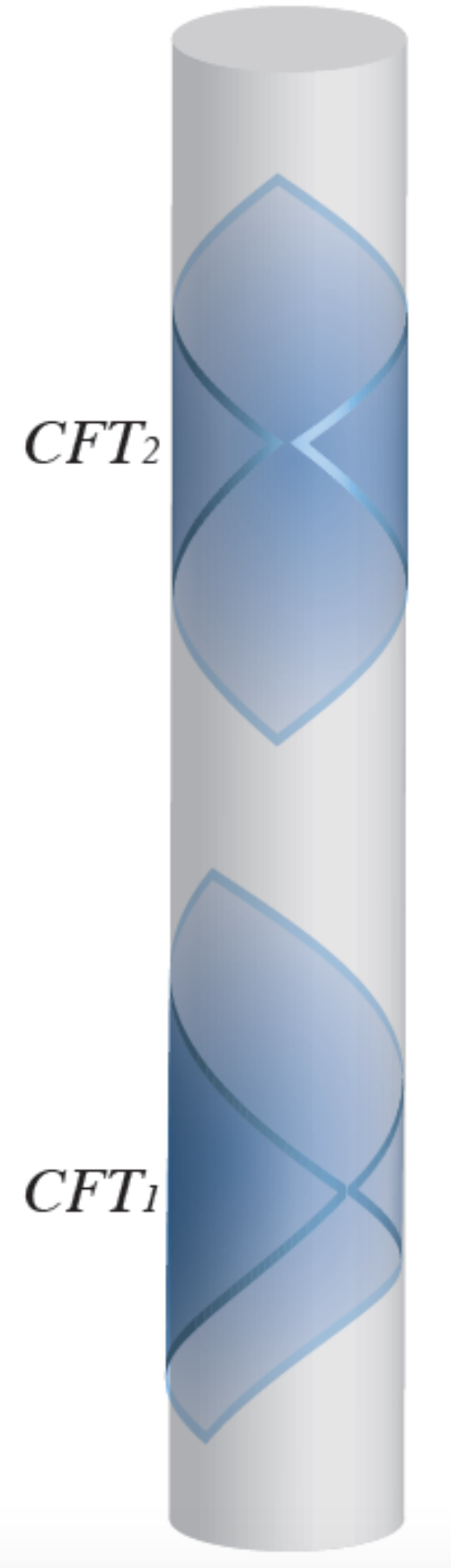}
\label{subfig:DoubleEinstein}
}
\hspace{1cm}
\subfigure[]{
\includegraphics[height=9cm]{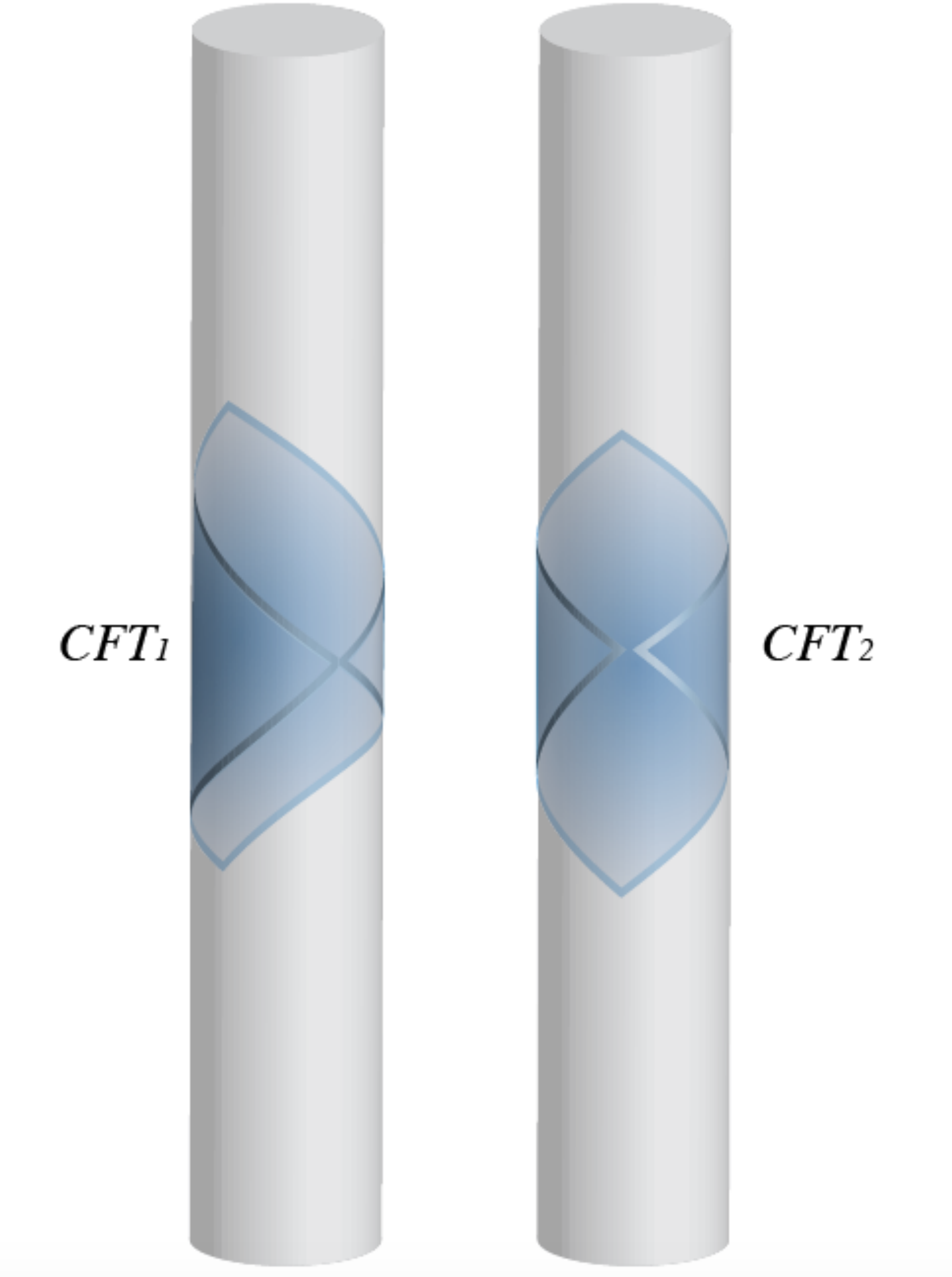}
\label{subfig:SeparateEinstein}
}
\caption{Two copies of a CFT can be mapped to \subref{subfig:DoubleEinstein} disjoint subsets of the same CFT in the Einstein static universe if they are dependent, or  \subref{subfig:SeparateEinstein} two separate Einstein static universes if they are independent.}
\label{fig:Einstein}
\end{figure}

In the context of holography, we will often be interested in quantum field theories that are  conformally invariant, so we will give a few examples of dependent and independent CFTs. The simplest example can be constructed from two copies of $\mathcal{N}=4$ super Yang-Mills (SYM) in the ground state on (3+1)-dimensional Minkowski space.  As is well known, each copy is holographically dual to a Poincar\'e patch of pure AdS$_{5}$.  Explicitly, the Minkowski metric  can be written as a conformal rescaling of part of the Einstein static universe: 
\begin{equation} \label{rescaling}ds^{2}_{flat} = \left [  \sec\left ( \frac{T+R}{2}\right)\sec\left ( \frac{T-R}{2}\right)\right]^{2} \left (-dT^{2} +dR^{2} +\sin^{2} R d\Omega^{2}\right).
\end{equation}
where $-\pi<T\pm R < \pi$ and $R>0$.  Thus each copy of $\mathcal{N}=4$ SYM on Minkowski space is a subset of $\mathcal{N}=4$ SYM on the Einstein static universe. The two copies are dependent if they share a Hilbert space, which can only happen if the above conformal transformation maps them onto the \textit{same} $\mathcal{N}=4$ SYM on the Einstein static universe. They are independent if it maps them to two \textit{separate} $\mathcal{N}=4$ SYM on two distinct copies of the Einstein static universe.  In the former case, the two copies of $\mathcal{N}=4$ SYM on Minkowski space are holographically dual to two Poincar\'e patches of the same global AdS spacetime (see Fig.~\ref{subfig:DoubleEinstein}). These CFTs can communicate both through the boundary and through the bulk. In the latter case (see Fig.~\ref{subfig:SeparateEinstein}), they are each dual to a Poincar\'e patch of a different global AdS spacetime: the CFTs cannot communicate in any way, either through the bulk or the boundary, in agreement with the NTP.  

This construction is not limited to Minkowski space: de Sitter (dS), for example, is also conformally related to a finite patch of the Einstein static universe, so
two  CFTs living on two copies of dS, or one copy on dS and the other on Minkowski space, can all in principle be dependent 
 -- that is, as long as the two copies of the CFT can be consistently mapped to the  same Einstein static universe. 

It is worth noting that any two nonconformal quantum field theories that live on separate, maximally extended spacetimes,  must be independent.  Conversely, as a caution we note that simply  taking any two finite subsets of a  CFT and conformally rescaling them into two complete universes does not always yield dependent CFTs. The problem is the following: consider a Cauchy surface in the original CFT and two open subsets $S_1$ and $S_2$.  Any pure state defined in one of the subsets will be singular in the larger CFT, since there are no correlations across the boundary of $S_i$ \cite{CzeKar12b}.  For example, the quantum stress tensor will diverge on this boundary. To avoid this problem, dependent CFTs must each contain an entire Cauchy surface of the larger CFT (or all but a set of measure zero, as in the case of Minkowski spacetime).

\begin{figure}[t]
\centering
\includegraphics[width=5cm]{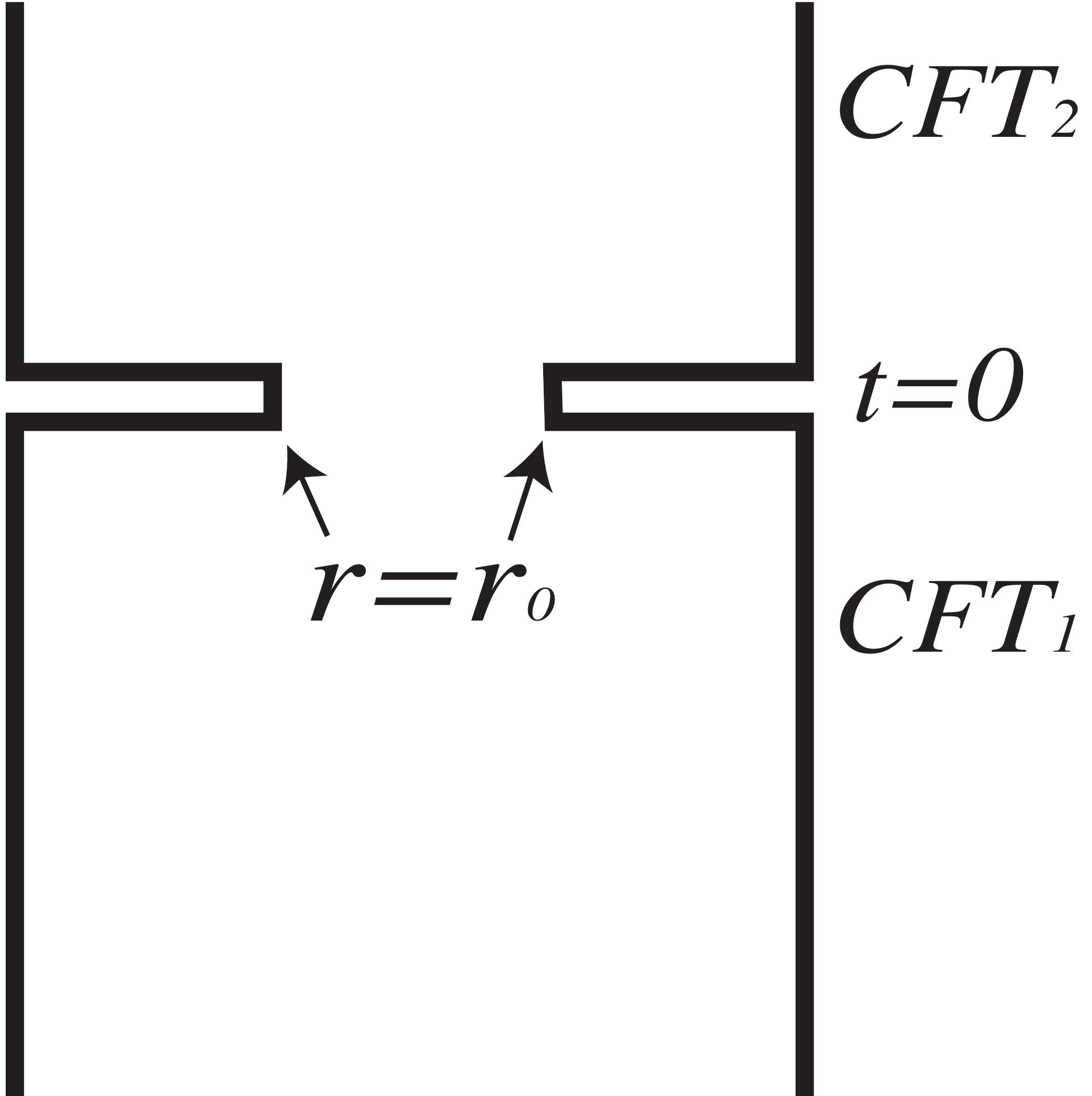}
\caption{By excising a constant time ($t=0$, $r>r_{0})$ surface from global AdS, we may artificially obtain a setup which a priori seems to have two independent CFTs, and yet bulk signals from CFT$_{1}$ to CFT$_{2}$ are allowed. This is not in violation of the No Transmission Principle, since the  two CFTs are simply subsets of the CFT that lives on the complete Einstein static universe.}
\label{fig:excised}
\end{figure}

As a final, somewhat pathological example, consider a holographic field theory on the Einstein static universe dual to global AdS, and excise the region $r > r_0 $ from the constant time slice $t=0$ in the bulk geometry. The result naturally picks out two dual field theories on two separate universes: the field theory that lives on $t<0$, and the field theory that lives on $t>0$ (see Fig~\ref{fig:excised}). Information about an excitation in the $t<0$ field theory can travel through the bulk spacetime to the $t>0$ field theory. This, however, is not in violation of the NTP: the two field theories in question are dependent --- they are subsets in this case of a field theory that lives on the complete Einstein static universe, with the $t=0$ slice not excised.

While these examples show that there exist dependent field theories, the class of pairs of field theories that are independent is by far larger.  We now turn to deriving constraints on the bulk spacetime from the No Transmission Principle.

\section{The No Transmission Principle For Singular Field Theories}\label{sec:singular}

A key question in the study of cosmological singularities is whether quantum gravity mediates a bounce from a big crunch. Is a cosmological singularity a true end to spacetime or is there another semiclassical region of spacetime before a big bang or after a big crunch? We will show in this section that a large class of holographic bounces are 
forbidden.\footnote{We define a bounce to be one in which signals can propagate from the past semiclassical spacetime to the future. If this is not possible, then operationally the spacetimes are not connected.}
For completeness, we will also discuss one case of a quantum bounce which is not yet ruled out. There is evidence that it is, but further work is needed to settle the question.

The crux of the argument is that the NTP implies that the bulk evolves forward in time only if the dual field theory evolves forward in time.  If the field theory evolution ends,  bulk evolution must stop; it is this principle which generally forbids bounces. A quantum field theory evolution will end only if it becomes singular in a suitable sense. For quantum field theories which are not  conformally invariant, the definition is simple: a field theory is singular if its evolution stops at some finite time. This can be a result of a curvature singularity in the underlying spacetime or a sickness in the field theory (e.g. a Hamiltonian which is unbounded from below).   We will review several ways in which this can happen, but first we must address the following issue: since many holographic setups involve CFTs, any finite time evolution can be rescaled to infinite time via a conformal mapping. We will therefore need a conformally-invariant way of defining arrested evolution. 

We consider CFTs on 
maximally conformally extended spacetimes, i.e., spacetimes which cannot be conformally embedded as a proper subset of a larger spacetime. We assume that the
 topology  is $X\times \mathbb{R}$, where $X$ is a compact $(d-1)$-dimensional manifold,\footnote{Spacetimes like Minkowski space that can  be conformally mapped to a compact universe are included.}
 and that the metric is globally hyperbolic: there exists a Cauchy surface ${\cal C}$ with unique classical evolution from initial data on ${\cal C}$ to anywhere in the spacetime. In fact, the spacetime can be foliated by a one parameter family of Cauchy surfaces $\{ {\cal C}_t\}$ each with topology $X$~\cite{Ger70}. Given a conformal metric on $X\times \mathbb{R}$, we pick a conformal frame in which the volume of ${\cal C}_t$ is bounded from above and below by nonzero constants for all $t$; such frames exist for any conformal metric (this excludes a conformal frame in which the CFT lives in a spacetime like de Sitter). We will call this a {\it standard conformal frame}.  The notion that evolution ends in finite time in a standard conformal frame is now conformally invariant.  \\

\noindent  \textit{{\bf Definition:} If the evolution of the CFT extends to infinite future proper time (in a standard conformal frame), then the CFT is {\it future-complete}. If in this frame, evolution cannot be extended to infinite proper time, the CFT is {\it future-singular}. Similarly, if there is past-infinite evolution of the CFT, the CFT is {\it past-complete}, and {\it past-singular} otherwise.}\\

Three different types of singular CFTs have been discussed in the literature: (1) the CFT could have a potential which is unbounded from below so observables run off to infinity in finite time, (2) the CFT could live on a spacetime with a cosmological singularity in the standard conformal frame (see Sec.~\ref{sec:GR} for a precise definition), (3) the CFT could be coupled to a time dependent source for a relevant deformation which becomes singular in a finite time. We now briefly review examples that illustrate these three possibilities. In each case the dual bulk geometry has a cosmological singularity.

 An example of the first case is given in  \cite{HerHor04, HerHor05} where a stable $(2+1)$-dimensional CFT is deformed by a marginal triple trace term: $S = S_0 + k\int {\cal O}^3$ where ${\cal O}$ is a dimension one operator. This adds a potential which is unbounded from below. A similar example is studied in \cite{CraHer07}, where a $(3+1)$-dimensional CFT 
 is deformed by a marginal double trace term: $S = S_0 + k\int {\cal \tilde O}^2$ where ${\cal \tilde O}$ is a dimension two operator. Since $k>0$, this again corresponds to adding a potential which is unbounded from below. In both cases, the bulk dual consists of gravity coupled to a scalar field $\phi$ with potential $V(\phi)$ coming from a consistent truncation of supergravity. It was shown that there are solutions in which smooth asymptotically AdS initial data evolves into a cosmological singularity in the bulk. 

Examples of the second possibility, that the field theory spacetime is itself singular, were discussed in  \cite{DasMich06, EngHer14, EngHer15}  where the CFT was defined on the Kasner universe:
\begin{equation} \label{Kasner}
ds^{2} = -dt^{2} +\sum\limits_{i=1}^{d-1} t^{2p_{i}}dx_{i}^{2},
\end{equation}
where $\sum p_i = \sum p_i^2 = 1$ and the $x_{i}$'s are periodically identified (to make space compact). This metric describes an anisotropic, homogeneous cosmology with a curvature singularity at $t=0$. This metric is not in a standard conformal frame (the volume of spatial slices vanishes at $t=0$), but is conformally related to a standard conformal frame by the conformal factor $t^{-1/(d-1)}$. There is a simple bulk dual to this setup, with the metric:
\begin{equation} 
ds^{2} = \frac{1}{z^{2}}\left(dz^{2} -dt^{2} +\sum\limits_{i=1}^{d-1} t^{2p_{i}}dx_{i}^{2}\right),
\end{equation}
which solves the vacuum Einstein equations with a negative cosmological constant. This bulk metric has a cosmological singularity at $t=0$ for all $z$, as well as a singularity at the Poincare horizon, $z = \infty$. Should the additional singularity be of concern, there is a related solution, the Kasner-AdS Soliton, which is also dual to a field theory on Eq. \eqref{Kasner} (with one of the $p_i=0$)~\cite{EngHor13}
\begin{equation}  ds_{\text{soliton}}^{2} = \frac{1}{z^{2}} \left [ \left(1-z^{d-1}\right) d\theta^{2}  -dt^{2} +\sum\limits_{i=1}^{d-1} t^{2p_{i}}dx_{i}^{2} + \frac{dz^{2}}{1-z^{d-1}} \right]. \label{SolitonMetric}\end{equation}
With a suitable period for $\theta$, space now smoothly caps off at $z=1$. 

Finally, we give an example of singular field theories in which the singularity stems from a singular time-dependent source~\cite{Mal10, HarSus10, BarRab11}.  A relevant perturbation of a CFT on de Sitter (with a constant coefficient of appropriate sign) is dual to a bulk with matter coupled to gravity\footnote{For some deformations, this is dual to gravity coupled to the same scalar fields used in the first example above, and the bulk solutions are in fact identical to the ones found in  \cite{HerHor04, HerHor05, CraHer07}.}. Within the future lightcone of a point in 
the bulk, the solution is described by an open FRW universe with a big crunch extending all the way through the bulk. Similarly, within the past lightcone, there is an expanding open FRW universe with an initial big bang extending all through the bulk. For small coefficients, the deformed CFT on de Sitter space is perfectly well-defined, but if one now conformally maps de Sitter to a standard conformal frame like the Einstein static universe, the coefficient of the relevant operator gets multiplied by a power of the conformal factor and  diverges in finite time both in the past and the future: the CFT is  being driven by a source that becomes singular~\cite{BarRab11}.

We will establish a correspondence between singular CFTs and singular bulks in the sections below. We first consider bounces in a bulk described by classical general relativity, then in a special case of classical string theory, and finally in full quantum gravity.

\subsection{In Classical General Relativity}\label{sec:GR}

We  argue below that a singular holographic field theory is dual to an asymptotically AdS spacetime with a cosmological singularity -- a singularity which extends out to infinity and cuts off all further evolution.  More precisely, let $\Sigma$ be a complete spacelike surface in the bulk, and let $\tilde g_{AB} = \Omega^2 g_{AB}$ be the rescaled bulk metric with timelike boundary at infinity attached.
\\

\noindent \textit{ {\bf Definition:}  The bulk spacetime has a future cosmological singularity (big crunch) if it is maximally extended and the length of all future directed timelike curves from  $\Sigma$ is bounded when computed with the rescaled metric $\tilde g_{AB}$. }\\

\noindent A big bang singularity is defined similarly with ``future" replaced by ``past". 

One can characterize the region of the bulk described by the dual field theory as follows: given a conformal boundary $B$ to an asymptotically AdS spacetime, the dual field theory on $B$ describes the region of the bulk spanned by all  spacelike surfaces which end on $B$.

In the particular case where a holographic field theory is singular due to a cosmological singularity on the boundary, we obtain a purely geometric result which relates the conformal geometry of the boundary to the geometry of the bulk. To state it formally, we first define what we mean by a boundary cosmological singularity:\\

\noindent \textit{ {\bf Definition:}  The conformal boundary $B$ of an asymptotically AdS spacetime has a  future cosmological singularity (a big crunch)  if  in any standard conformal frame, through every point of any Cauchy surface ${\mathcal{C}_t}$, there exists a future incomplete timelike geodesic.}\\

\noindent  Once again, a big bang singularity is defined similarly with ``future" replaced by ``past". Note that we cannot require that the length of all timelike curves (or even all timelike geodesics) be bounded in this case, since we can rescale the metric by a function that diverges locally near the singularity. We now state a purely geometric implication of the NTP:\\

\noindent {\bf Conclusion:} {\it Consider any supergravity theory arising in the low energy limit of string theory.  If the conformal boundary metric of an asymptotically AdS solution has a cosmological singularity, then the bulk solution must also have a cosmological singularity. }\\

\begin{figure}[t]
\centering
\subfigure[]{
\includegraphics[width=5cm]{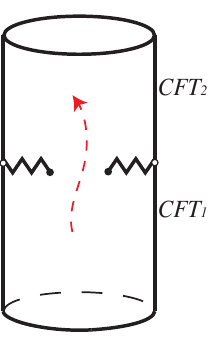}
\label{subfig:WithHole}
}
\hspace{1cm}
\subfigure[]{
\includegraphics[width=5cm]{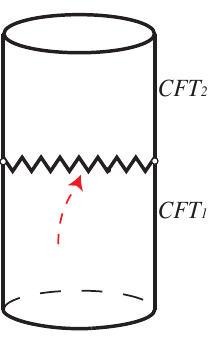}
\label{subfig:WithoutHole}
}
\caption{\subref{subfig:WithHole} In a bulk in which a singularity that extends to the boundary has a hole, signals can travel from the bulk region dual to CFT$_{1}$ to the bulk region dual to CFT$_{2}$. This is forbidden by the No Transmission Principle, as CFT$_{1}$ and CFT$_{2}$ are independent.  \subref{subfig:WithoutHole} The NTP implies that any signal in the bulk region dual to CFT$_{1}$ must terminate before it reaches the bulk region dual to CFT$_{2}$: the singularity has no hole.}
\label{fig:SingularityWithHole}
\end{figure}

\noindent It is easy to see that this follows immediately from the NTP. If the conformal boundary has a cosmological singularity in the future (as defined above), evolution of the holographic field theory dual must end in finite time. This is the second type of singular field theory discussed in Sec.~\ref{sec:singular}. If the bulk dual  does not have a big crunch singularity, then evolution in the bulk can continue into another region of spacetime with its own asymptotic boundary metric. This would require that the bulk be dual to two field theories: the original future-singular field theory, and another, past-singular field theory which can receive causal signals via the bulk from the future-singular field theory. See Fig.~\ref{subfig:WithHole}. The two field theories are manifestly independent, so any communication between them violates the NTP. This contradiction is avoided only if the bulk has a cosmological singularity, as in Fig~\ref{subfig:WithoutHole}, establishing the above result.

The above implication is a novel singularity ``theorem'': a boundary singularity guarantees a bulk singularity. Since it is purely a geometric statement, it is reasonable to expect that it can be shown just using techniques from classical general relativity. Ideally, the restriction to supergravity theories would be replaced by a more general condition like the Null Energy Condition ( $T_{ab} \ell^a \ell^b \ge 0$ for all null $\ell^a$). Since the result shows that the singularity cannot end in a naked singularity in the interior as illustrated in Fig.~\ref{fig:SingularityWithHole}, it also incorporates a (very limited) form of cosmic censorship.

The above conclusion is not obvious: there are bulk spacetimes which are nonsingular, but still have a conformal boundary with a cosmological singularity.  As argued above, the NTP shows that these are not solutions to string theory. As an example, consider the following spacetime:
\be\label{reg}
ds^2 = \frac{1}{z^2} \left [  -dt^2 + \sum_{i=1}^{d-1} (t^2 + z^2)^{p_i} dx_i^2 + dz^2\right ]
\ee
where $\sum p_i = 0$. This form of the metric is intended to apply only in the asymptotic (small $z$) region. (For large $z$, it can be modified in any nonsingular manner.)
The bulk spacetime given by \eqref{reg} is suggestive of a regulated version of the AdS Kasner metric in which the cosmological singularity is replaced by a bounce. The metric on the boundary $z=0$ is still singular and is simply Kasner \eqref{Kasner} in a standard conformal frame. The NTP, as argued above, shows that this metric is forbidden as a solution to general relativity with matter coming from string theory. Since this spacetime violates the Null Energy Condition, the results of the NTP agree with conventional wisdom, that any reasonable classical geometry arising from string theory should obey the Null Energy Condition.

We have focused above on the case in which the field theory has a cosmological singularity, but the conclusion holds whenever the field theory is singular, irrespective of the type of singularity: a portion of a bulk spacetime which is dual to a future-singular field theory can have no overlap with a portion of the bulk which is dual to a past-singular field theory.

\subsection{In Classical String Theory}\label{sec:ST}

In this section, we explore one implication of the No Transmission Principle for holography with infinite $N$ and finite coupling, dual to a classical, but stringy bulk. Can classical stringy effects mediate evolution through the singularity? Since the argument in the previous section made no use of $N$ or the 't Hooft coupling, we conclude that the same line of reasoning shows that the answer is no. 

We illustrate this with an example of one classical stringy effect that can be explicitly ruled out as mediating evolution through a singularity: tachyon condensation. When one direction of space is compactified to a circle, string theory has states corresponding to winding modes around the circle. These winding modes become tachyonic whenever the size of the circle is smaller than the string scale and fermions are anti-periodic around this circle. A space that has a circle at infinity which is larger than the string scale and slowly shrinks to become smaller than the string scale in the interior has a tachyon instability; this instability results in the circle pinching off~\cite{AdaLiu05}. 

\begin{figure}[t]
\centering
\subfigure[]{
\includegraphics[width=6.5cm]{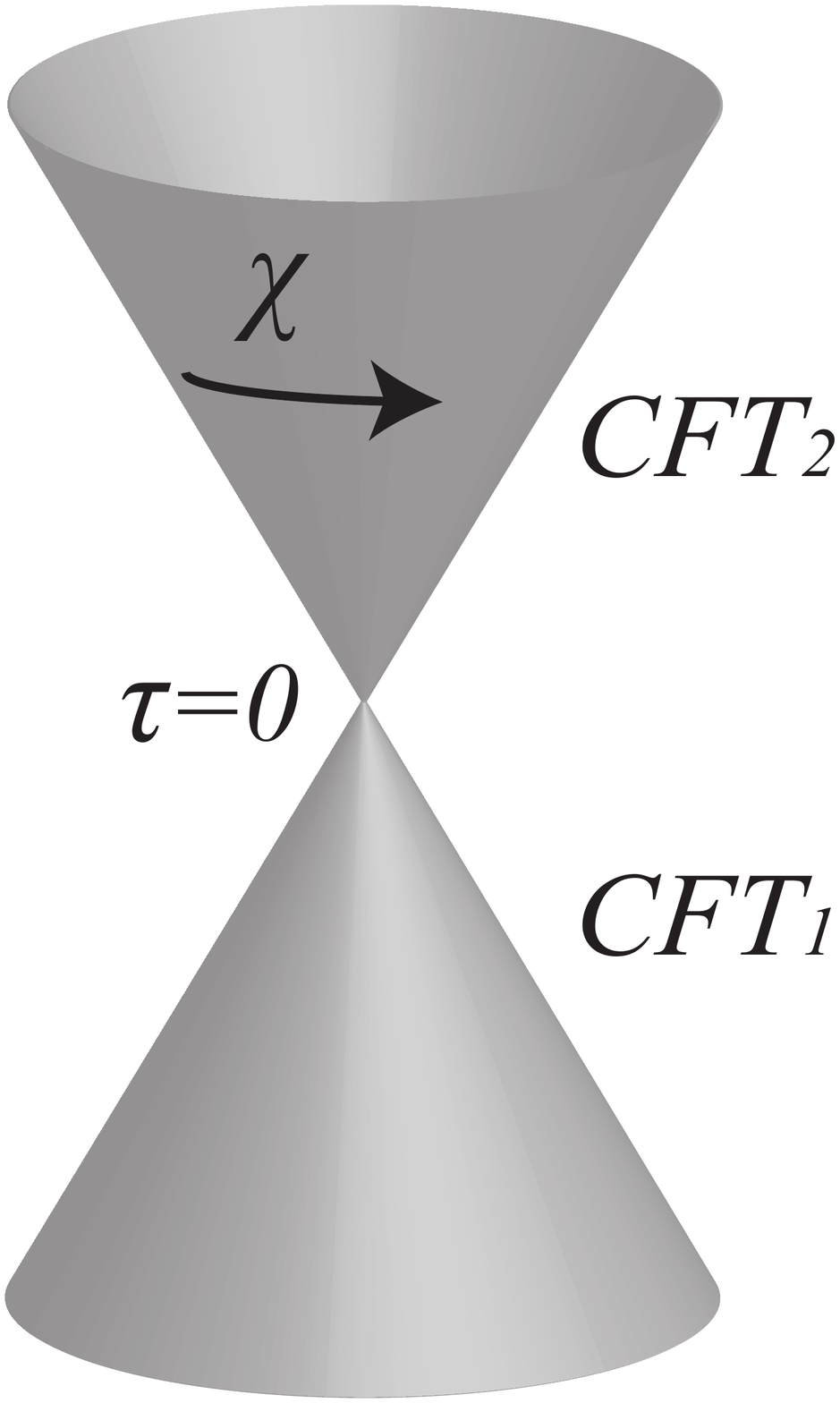}
\label{subfig:Milne}
}
\hspace{1cm}
\subfigure[]{
\includegraphics[width=5cm]{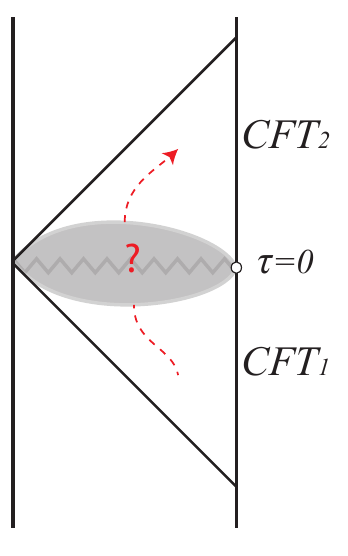}
\label{subfig:TachyonTunneling}
}
\caption{\subref{subfig:Milne} Two independent CFTs on the past and future Lorentzian cone, with a conical singularity in between them.  \subref{subfig:TachyonTunneling} The bulk dual with a tachyon condensate modifying the near-singularity region. It may a priori seem that the tachyon could 
allow signals to propagate through, but the No Transmission Principle forbids the signal in the shaded region from emerging in the bulk region dual to CFT$_{2}$.}
\label{fig:tachyon}
\end{figure}

In the context of cosmological singularities, the tachyon instability can manifest as a result of the circle shrinking in time. An explicit instance of this was worked out in~\cite{McGSil05} for the Milne spacetime~\cite{Mil35}:
\be
 ds^2 = -d\tau^2 + \tau^2 d\chi^2.
 \ee
  These coordinates cover the interior of the light cone of two dimensional Minkowski space. Compactifying $\chi$ with period $\chi_0$ forms a Lorentzian cone, as illustrated in Fig.~\ref{subfig:Milne}. Since the length of the circle is $L(\tau)= |\tau|\chi_0$, its rate of change is $\dot L =\pm \chi_0$.  It was shown in \cite{McGSil05}  that if $\chi_0$ is sufficiently small, the string spectrum can be reliably computed and a winding string mode becomes tachyonic when $L$ reaches the string scale\footnote{We are ignoring the extra spatial dimensions which will play no role in this discussion.}. When the tachyon condenses, spacetime ends. All string modes (including the graviton) become massive and there is no low energy spacetime description of the physics \cite{McGSil05}. This is a classical string theory effect which does not involve quantum gravity. 
  
  The  above metric with $-\infty < \tau < +\infty$  describes two cones attached at a singular tip (Fig.~\ref{subfig:Milne}). Although the tip is only a conical singularity (since the curvature vanishes for all nonzero $\tau$), in classical string theory it is unstable to developing large curvature: the slightest perturbation with momentum around the circle will become strongly blue-shifted near $\tau = 0$ and cause the curvature to become large. The NTP shows that, at least in the context of holography, the tachyon instability (which can set in before the curvature becomes large) will not allow signals  
 to propagate
 from the past cone to the future cone through the tachyon condensate.

More explicitly, consider $AdS_3$ in Poincare coordinates and write the Minkowski metric on each radial slice in Milne coordinates: 
\be
ds^2 = \frac{1}{z^2}\left [-d\tau^2 + \tau^2 d\chi^2 + dz^2\right ] 
\ee
where $\chi$ is compactified with period $\chi_0$. The bulk spacetime has a cosmological singularity at $\tau = 0$: this is a simple example of an AdS cosmology. The dual field theory consists of two independent field theories on two Lorentzian cones which are attached at their tip. In a standard conformal frame, the past cone by itself is conformal to an infinite cylinder: $ds^2 = -dt^2 + d\chi^2$ with $t = \ln \tau$; the field theory on the past cone itself is nonsingular. In the bulk, tachyon condensation replaces the singularity at $\tau = 0$ with a tachyon condensate when $|\tau| L/z < \ell_s$, see Fig~\ref{subfig:TachyonTunneling}.  If signals in the bulk could propagate through the tachyon condensate, it would violate the NTP. Note that although the tachyon condensate extends out to infinity, it is confined to a region $|\tau | < z\ell_s/L$. At the conformal boundary, then, it is concentrated at the tip and does not affect the geometry of the cone, or the infinite static cylinder.

\subsection{In Quantum Gravity}\label{sec:QG}

A holographic field theory at finite $N$ and finite coupling is dual to a bulk spacetime with all quantum string theory effects included. If the dual field theory is  singular, the No Transmission Principle implies that signals cannot be transmitted through cosmological singularities -- even in a quantum stringy bulk. In other words, such bounces are forbidden. A potential objection to this argument is that holography might fail in this context. After all, gravitational excitations are no longer confined inside a finite box. Since quantum gravity effects are strong near the singularity, and the singularity extends all the way out to infinity, perhaps decoupling fails and the boundary metric will be subject to quantum fluctuations\footnote{We thank S. Hartnoll for raising this point.}, rather than being a fixed, nondynamical background as usually assumed. 
  
The following argument shows that holography does not fail, and can still be applied in the cosmological context.  
First we show that the boundary metric (in Fefferman-Graham gauge) is singular for any  semiclassical bulk geometry with large asymptotic curvature, even if the bulk curvature remains finite. In other words, bulk quantum effects cannot regulate the boundary metric so that it becomes nonsingular. The line of reasoning is simple: no matter how large 
the boundary curvature is, as long as it remains finite, the bulk remains asymptotically AdS everywhere. In particular, the asymptotic bulk curvature is set by the bulk cosmological constant and does {\it not} become large. To see this, recall that on any scale smaller than the boundary curvature, the boundary metric looks flat, so the bulk metric resembles standard AdS in that asymptotic region. 

More explicitly, for any nonsingular $d$-dimensional boundary metric $g^0_{\mu\nu}$, introduce Fefferman-Graham coordinates in which the bulk metric takes the asymptotic form 
\be
ds^2 = \frac{1}{z^2} [dz^2 + g^0_{\mu\nu} dx^\mu dx^\nu].
\ee
The scalar curvature of this metric is $R = -d(d+1) +z^2 R_0$ where $R_0$ is the scalar curvature of $g^0_{\mu\nu}$. So as long as $R_0$ remains finite, it yields only a subleading contribution to the asymptotic bulk curvature. (A similar result holds for the full Riemann tensor.) This implies that if the bulk metric has large finite curvature that continues into the asymptotic region (as expected for any quantum backreaction near a cosmological singularity), the boundary metric must be singular.  

Quantum gravity effects are expected to be significant only within a bounded distance to the singularity. Suppose  that the singularity is at $ t= 0$ and the quantum effects are significant within a proper time $\tau_0$ of the singularity. Applying a conformal rescaling which produces a finite boundary metric, we find that all effects within a proper time $\tau_0$ of the asymptotic singularity are shrunk to zero proper time on the boundary. In other words, they are compressed to the $ t=0$ singularity in the boundary metric, and the boundary metric at any nonzero time does not fluctuate. This can also be seen by conformally rescaling any singular boundary metric to a (de Sitter-like) frame in which the singularity is pushed off to infinity. In this frame, the curvature on the boundary remains bounded, and the standard rules of holography should apply. Quantum gravity effects in the bulk should not affect this boundary metric. A rescaling back to the singular boundary metric shows that the boundary metric away from the singularity remains fixed.

Thus, the basic rules of holography still apply. Our conclusion from the NTP holds: a singular field theory cannot be holographically dual to a cosmological bounce in full quantum string theory.

\subsection{Can a Big Crunch be Dual to a Nonsingular Field Theory?}\label{sec:bounce}

We are aware of only one construction in which the bulk spacetime has a cosmological singularity, but it is not immediately obvious that the boundary field theory is singular. Consider $\mathcal{N}=4$ super Yang-Mills theory on four-dimensional Minkowski space with a time dependent coupling $g_{YM}(t)$ that vanishes at $t=0$ \cite{AwaDas08}; this is an example of a marginal coupling which is taken to extreme values. The bulk dual consists of gravity coupled to a massless scalar field (the dilaton). Since $g_{YM}$ determines the asymptotic value of the dilaton via $e^\phi = g_{YM}^2$, the vanishing of the coupling requires that $\phi(t) \to -\infty$ as $t \to 0$. The backreaction of this diverging scalar field causes a cosmological singularity in the bulk. For example, one of the bulk solutions that has been found is
\be
ds^2 = \frac{1}{z^2}[ dz^2 + |t|\eta_{\mu\nu}dx^\mu dx^\nu], \qquad e^{\phi(t)} = g_s|t|^{\sqrt 3}
\ee 
which has a cosmological singularity at $t=0$.\footnote{Note that the boundary metric in Fefferman-Graham coordinates is singular, consistent with the general result discussed above. But since the singularity is entirely in the conformal factor, one can view the super Yang-Mills theory as living in Minkowski spacetime.} In the field theory, it is not obvious that evolution must stop. Indeed the vanishing of the coupling might suggest that the theory just becomes free near $t=0$. It was shown in \cite{AwaDas08} that this is not the case. The fields become large near $t=0$ and the interactions never become negligible. Furthermore, they present strong evidence that evolution must end. The state appears to become singular with diverging energy and a wildly oscillating phase as $t\to 0$. This is not yet a proof that evolution must end since effects of renormalization were not fully taken into account. If it could be shown that (at finite $N$ and coupling) there is some choice of $g_{YM}(t)$ such that evolution is well defined through a time when the coupling vanishes, then evolution in the bulk must continue as well.  This would be the first example of a holographic bounce. Note that this is not in contradiction with the above discussion, which assumed that the dual field theory was singular. 

If  $g_{YM}(t)$ becomes very small but stays bounded away from zero, then the field theory is clearly nonsingular and evolution can continue for all time. This, however,  tells us nothing about cosmological bounces: the bulk  solution now contains a large black hole \cite{AwaDas08}, and the singularity does not enter the asymptotic region. The fact that the field theory evolution  continues for all time simply reflects future-infinite evolution outside of the black hole.  This is very similar to our discussion in the previous subsection of  regulating the boundary metric so that its curvature remains finite. Both types of modifications destroy the cosmological singularity.

\section{The No Transmission Principle for Nonsingular Field Theories}\label{sec:nonsingular}

In this section we discuss consequences of the No Transmission Principle for nonsingular holographic field theories. We find the exclusion of the following two a priori possible scenarios: (i) evolution through black hole singularities
to a new asymptotically AdS region of spacetime, (ii) existence of traversable wormholes connecting two asymptotic regions.
 Here we will assume that the CFT and its superselection sectors contain complete information about the black hole interior. We discuss the alternative in Sec.~\ref{sec:discussion}.

\begin{figure}[t]
\centering
\includegraphics[width=5cm]{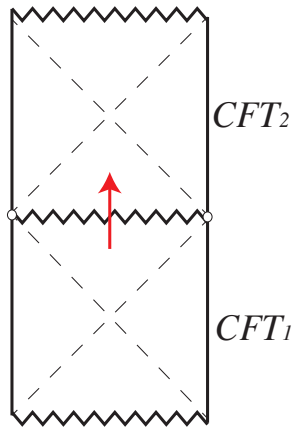}
\caption{Resolution of the Schwarzschild black hole singularity allowing evolution to another asymptotic region would violate the No Transmission Principle.}
\label{fig:SchwarzschildEvolution}
\end{figure}

\vskip 1cm
\noindent \textbf{No Evolution through Black Hole Singularities}: 
Could a signal entering an AdS black hole pass through the black hole and reemerge in another asymptotically AdS spacetime? In the simplest case of 
a Schwarzschild black hole in global AdS, it is a priori possible that quantum gravity could mediate evolution through the singularity into another asymptotic region (see Fig.~\ref{fig:SchwarzschildEvolution}). This kind of evolution, however would result in 
signals propagating from the field theory dual to the black hole to the field theory dual to the white hole. Since the two field theories
each live on a complete Einstein static universe, they are independent, and this is immediately forbidden by the NTP. We find that in \textit{any} regime of string theory, the black hole singularity in Schwarzschild-AdS does not admit evolution to another asymptotic region.

\begin{figure}[t]
\centering
\includegraphics[width=7cm]{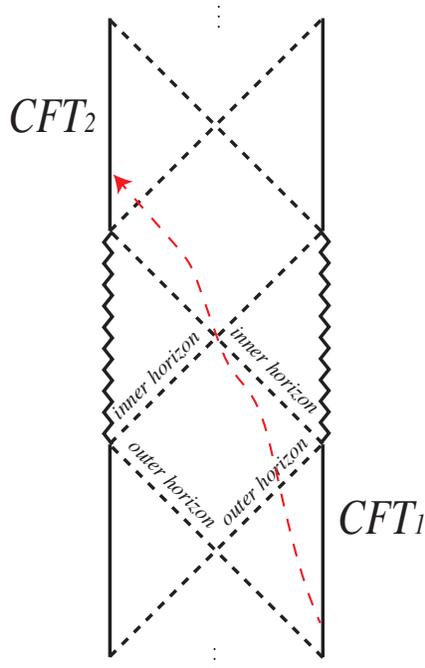}
\caption{A conformal diagram of a charged black hole, with the possible path of a signal between independent CFTs. Any such signal, however, collapses the inner horizon into a null singularity. The No Transmission Principle implies that there is no evolution past this null singularity.}
\label{fig:RNAdS}
\end{figure}

Generic black holes, however, have some nonzero rotation (and possibly nonzero charge). The resulting causal structure appears at first to facilitate the transmission of 
signals between asymptotic regions even at the classical level. The Penrose diagram for a charged and rotating AdS black hole contains multiple asymptotic regions (see Fig.~\ref{fig:RNAdS}). Each asymptotic region is dual to a CFT on a complete Einstein static universe, and each of the dual CFTs is clearly independent of all of the others. It may prima facie appear that this system presents a violation of the NTP: signals can travel from one asymptotic region to another through the bulk, as illustrated in Fig.~\ref{fig:RNAdS}. These signals result in a transmission of information which cannot be explained by entanglement and is of course forbidden by the NTP. This puzzle is resolved by the instability of the inner horizon of such black holes: any excitation that enters the black hole will cause the inner horizon to develop a curvature singularity~\cite{PoiIsr90, SimPen73, Ori91, Daf04}. The reader may raise an objection at this point: normally, a field theory excitation with energy less than $N^{p}$ for some suitable power $p$ does not result in any backreaction on the bulk metric in the large $N$ limit. This argument, however, assumes bulk stability. In the presence of an instability, the curvature diverges for any excitation at finite $N$, so the natural large $N$ limit contains a singularity for any excitation\footnote{We thank Don Marolf for a discussion on this point.}. 

The nature of the so-called weak null singularity that forms on the inner horizon has raised speculation regarding possible evolution beyond it. While the Christoffel symbols at the singularity are not square integrable, the metric at the singularity remains continuous. Since the total tidal distortion remains finite across the singularity, there has been some debate on the possibility that signals could pass through the singularity even in the purely classical regime. We claim that this cannot happen: such a scenario would result in the exchange of 
signals between independent field theories, and is therefore ruled out by the NTP. Most significantly, evolution through the weak null singularity to another asymptotic region is ruled out by the NTP even in full quantum gravity.

Finally, we comment on the asymptotic regions of an eternal AdS black hole which are not causally connected, where the maximally extended spacetime is not a pure state of a single field theory. The above argument is unaffected by these additional asymptotic regions: given any entangled state, an excitation can always be added to one field theory and allowed to propagate. 
Alternatively, one can combine the two field theories dual to the asymptotic regions on either side of the Einstein-Rosen bridge into one Hilbert space, and apply the NTP to these larger Hilbert spaces. 
This minor complication may at any rate be avoided via the formation of black holes from the collapse of a charged spherical shell. Such systems initially have only a single asymptotic region, and can be dual to a field theory in a pure state. If the local charge density of the shell is greater than its local mass density, then the shell will reach a minimum radius inside the horizon and bounce out into another asymptotic region (provided it is given sufficient kinetic energy at collapse)~\cite{Bou73}. With AdS boundary conditions, the shell will  continue to oscillate and execute an infinite number of bounces. The spacetime outside the shell is exactly Reissner-Nordstrom-AdS, and the inner horizon is again unstable. So this provides a pure state, or one sided, version of the argument that there is no evolution through inner horizon singularities.

Of course small black holes can evaporate in quantum gravity, and the fact that the evolution must be unitary already provided  an indication that no information could pass through the black hole into another universe.

\begin{figure}[t]
\centering
\includegraphics[width=10cm]{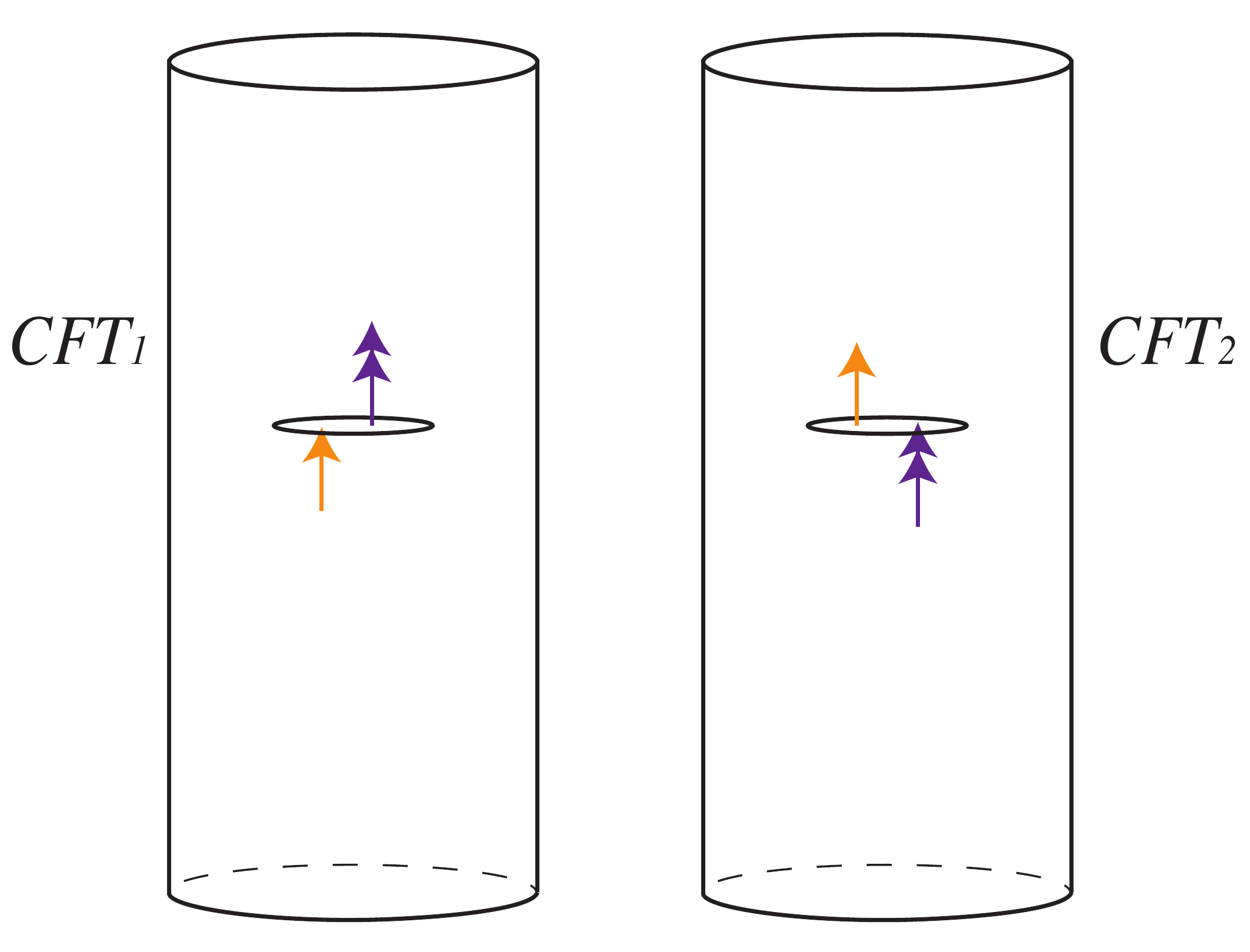}
\caption{A traversable wormhole obtained by identifying a spatial surface in two separate AdS universes. The orange signal (single-arrow), initially dual to some excitation in CFT$_{1}$, propagates into CFT$_{2}$; the purple signal (double-arrow), initially dual to an excitation in CFT$_{2}$, propagates from CFT$_{2}$ to CFT$_{1}$. This violates 
the No Transmission Principle.}
\label{fig:traversable}
\end{figure}

\vskip 1cm
\noindent \textbf{No Traversable Wormholes in Quantum Gravity}: A traversable wormhole may be constructed from two copies of global AdS, each with a dual CFT on $S^{d-1} \times \mathbb R$: excise a unit ball on a time slice $t=0$ from each spacetime and glue the bottom edge of one to the top edge of the other (see Fig.~\ref{fig:traversable}). A signal with $t < 0 $ in the first spacetime can now propagate to $t>0$ in the second spacetime.  This wormhole does not have negative energy or large curvature anywhere, but it has a naked singularity (defined in terms of geodesic incompleteness) along the boundary $r=1$. Could this singularity be smoothed out in quantum gravity? 
The answer is clearly no, since this wormhole violates the No Transmission Principle.

Now consider a permanent traversable wormhole: a static bulk spacetime with two (or more) asymptotically AdS regions each with a dual field theory\footnote{
The dual field theories will almost certainly be in some entangled state; however,  
signals cannot be transmitted via entanglement.}.  A bulk excitation near the first boundary can propagate to a second asymptotic region through the wormhole. Such an excitation initially corresponds to exciting a field theory dual to one asymptotic region; in this initial stage, the field theory on the second asymptotic region is ignorant of this excitation. Once the excitation propagates through the wormhole, however, it is identical to an excitation created by the second field theory, and the second field theory therefore must register the signal. Such communication is in clear violation of the NTP, and we therefore conclude that traversable wormholes connecting two asymptotic regions are forbidden, not just in classical or semiclassical gravity, but in full 
quantum gravity\footnote{A classical solution describing a traversable wormhole has recently been constructed using matter satisfying the null energy condition, but it involves closed timelike curves \cite{Ayon-Beato:2015eca}.}.

\section{Discussion}\label{sec:discussion}

We have deduced several nontrivial consequences of the simple statement that two quantum field theories cannot transmit signals to one another unless they share a common Hilbert space (and hence are both part of a larger field theory). It is rather surprising that a number of deep properties of full quantum gravity follow immediately from this simple statement.
Of course our conclusions are restricted to the theory of quantum gravity with asymptotically AdS boundary conditions obtained from gauge/gravity duality; other theories of quantum gravity might behave differently. 

In Sec.~\ref{sec:nonsingular} we assumed that the dual field theory contains complete information about the black hole interior. 
This is almost certainly true for eternal black holes with two asymptotic regions on an initial static surface:  it is generally agreed that a maximally entangled state of the two theories corresponding to a thermofield double describes the region inside as well as outside the horizon \cite{Mal01}. If the dual field theory does not describe the interior of a single-sided black hole, then we cannot rule out the  possibility that quantum gravity permits signals to propagate through black hole or cosmological singularities to a region inside another horizon.

As is standard in holography, 
we have been assuming that if an asymptotically AdS spacetime has several asymptotic regions, there is a separate dual field theory associated with each one. In particular, there is no coupling between them. If  such couplings did exist, the No Transmission Principle would clearly not apply.  Could our conclusions be avoided by simply extending holography to allow couplings between the dual theories? The answer is no. Most of our examples involve bulk spacetimes in which one asymptotic region is to the future of the other. In this case any coupling between the dual field theories would violate causality in the bulk, since they would allow  signals to be sent  into the past as well as the future. The one exception is the possibility of a traversable wormhole;  even this case, however,  cannot be described by any simple coupling between the dual theories, such as the product of a single trace operator in one theory with a single trace operator in the other. This type of  double trace  deformation corresponds to modified boundary conditions in the bulk and is more analogous to gluing the two boundaries together. 
A signal can pass from one theory to the other, but it does so instantly, without going through the wormhole. 

As mentioned earlier, the only way to avoid our conclusions is to add to holography some ad hoc rule that matches the future state of one dual theory with the past state of another theory on a disconnected spacetime. No one has found such a rule that works in all cases, and more importantly, there is some indication that such a rule does not exist.  Consider first the 
black hole examples discussed in Sec.~\ref{sec:nonsingular}. One would need to  evolve a state in one dual theory for an infinite time and extract some asymptotic limiting state to use in the next copy of the theory. However, a limiting state almost never exists.  The state can always be expanded in energy eigenstates, and unless the state is a single eigenstate, it will continue to evolve for all time. Certain observables might settle down, e.g., to their thermal expectation values, but there will still be fluctuations;  more importantly,  the microstate itself never settles down. Simply postulating an isomorphism between the Hilbert spaces and identifying all states of one dual theory with the other would violate causality: excitations in the future theory would then be seen in the past.

We have been implicitly using a Schr\"odinger representation in this discussion. In a Heisenberg picture, any evolution through the bulk black hole singularity would define something like an S-matrix between the Hilbert space of $CFT_1$ and that of $CFT_2$. This would contain nontrivial physics which is not captured by the dual field theory, contradicting the statement that the two dual descriptions are equivalent.

The ad hoc rule fairs no better with the singular field theories discussed in Sec.~\ref{sec:singular}. These theories are driven by external sources which diverge in finite time, or by the background metric itself becoming singular.
Since normal evolution ends in finite time, the limiting state is likely to be very badly behaved. Often the energy, and possibly other expectation values, diverge. (Indeed, a typical state of finite energy is expected to describe a black hole in the bulk.) It is unlikely that one can match such a singular end state with some initial state of another copy of the dual theory.\footnote{When the Hamiltonian is unbounded from below, one might try to construct a well defined operator via a self-adjoint extension. While this is straightforward in quantum mechanics, previous attempts to define it in quantum field theory do not lead to bouncing cosmologies \cite{HerHor05, CraHer07}, and it may be impossible to do so.}

We conclude by noting that our results support the idea of superselection sectors in holography~\cite{MarWal12}.   Given  CFTs on two disjoint copies of Minkowski spacetime,  local observables cannot distinguish  whether they share a Hilbert space or not; ancillary information is necessary to determine whether the bulk geometry is one copy of AdS or two.  This additional data is contained in
non-local observables such as two-point correlators or mutual information. In other words, the knowledge of whether the two CFTs are dependent or independent is not accessible to local observables. This is reminiscient of the superselection sectors of~\cite{MarWal12}, where it was shown that local observables of two entangled CFTs cannot determine whether the two CFTs are dual to a wormhole or two disconnected universes.

\end{spacing}
\section*{Acknowlegements}
It is a pleasure to thank W. Donnelly, D. Engelhardt, D. Jafferis, S. Fischetti, S. Hartnoll, 
J. Maldacena, D. Marolf, J. Polchinski, H. Reall, and A. Wall for discussions. 
This work began during the KITP program on Quantum Gravity Foundations: UV to IR, and GH thanks the KITP for its hospitality.
 This work was supported in part by NSF grant PHY-1504541. The work of NE was supported by the NSF Graduate Research Fellowship under grant DE-1144085 and by funds from the University of California. 

\bibliographystyle{JHEP}

\bibliography{all}

\end{document}